\documentclass[reprint,amsmath,amssymb,aps]{revtex4-1}

\usepackage{color}
\usepackage{amsmath}
\usepackage{graphicx}
\usepackage{dcolumn}
\usepackage{bm}
\usepackage{booktabs}
\usepackage{multirow}

\begin{document}
\preprint{APS/123-QED}

\title{Topological Protection of Two-photon Quantum Correlation on a Photonic Chip}

\author{Yao Wang,$^{1,2}$ Xiao-Ling Pang,$^{1,2}$ Yong-Heng Lu,$^{1,2}$ Jun Gao,$^{1,2}$ Zhi-Qiang Jiao,$^{1,2}$ Hao Tang,$^{1,2}$}
\author{Xian-Min Jin$^{1,2,}$}
\email{xianmin.jin@sjtu.edu.cn}

\affiliation{
	$^1$State Key Laboratory of Advanced Optical Communication Systems and Networks, School of Physics and Astronomy, Shanghai Jiao Tong University, Shanghai 200240, China\\
	$^2$Synergetic Innovation Center of Quantum Information and Quantum Physics, University of Science and Technology of China, Hefei, Anhui 230026, China}
\date{\today}

\maketitle

\textbf{Low-decoherence regime plays a key role in constructing multi-particle quantum systems and has therefore been constantly pursued in order to build quantum simulators and quantum computers in a scalable fashion. Quantum error correction and quantum topological computing have been proved being able to protect quantumness but haven't been experimentally realized yet. Recently, topological boundary states are found inherently stable and are capable of protecting physical fields from dissipation and disorder, which inspires the application of such a topological protection on quantum correlation. Here, we present an experimental demonstration of topological protection of two-photon quantum states on a photonic chip. By analyzing the quantum correlation of photons out from the topologically nontrivial boundary state, we obtain a high cross-correlation and a strong violation of Cauchy-Schwarz inequality up to 30 standard deviations. Our results, together with our integrated implementation, provide an alternative way of protecting quantumness, and may inspire many more explorations in `quantum topological photonics', a crossover between topological photonics and quantum information.}\\

\noindent Single photons inherently hold the features of single qubit in quantum computing, and has been widely used in various quantum simulation protocols, such as quantum walk~\cite{QW_1D,QW_2D}, boson sampling~\cite{BS_1,BS_2,BS_3,BS_4} and quantum fast hitting~\cite{FH}. The single-particle quantum walks have a precise mapping to classical wave phenomena. However, the advantage of uniquely quantum mechanical behavior is limited due to the single walker. 

In contrast, multiple indistinguishable particles can provide distinctly non-classical correlations, and this quantum behavior becomes a computational advantage. For example, two-particle quantum walk can be an algorithmic tool for the graph isomorphism problem~\cite{g2_QW}, and the universal computation can be achieved by multi-particle quantum walk efficiently~\cite{Q_multi}. Thus, it is crucial to preserve the non-classical features when constructing a multi-particle quantum computers.
 
Quantum error correction is proposed to preserve logical quantum states in a subspace and rectify errors according to the measurement outcomes of ancillary particles ~\cite{QEC_exp,QEC_rev}. Quantum topological computing, meanwhile, strives to store and manipulate quantum information with topological protection in a nonlocal manner using non-Abelian anyons~\cite{Top_Q_com}. Both of them are promising candidates in theoretical predictions but are still in their initial stage for experimental implementations~\cite{QEC_exp,QEC_rev,Top_Q_com,Top_Q_com_book}.

Topological photonics, derived from the discovery of topological phases in condensed-matter physics, aims to topologically protect photons from the inevitable fabrication-induced dissipation and disorder~\cite{Topo_review_1,Topo_review_2}. Many types of topological phases have been observed, for instance, Hall effect~\cite{,Topoto_hall_1,Topoto_hall_2}, edge states~\cite{Topo_edge_1,Topo_edge_2,Topo_edge_3,QC1}, topological insulators~\cite{Topoto_insulator_1,Topoto_insulator_2} and Weyl points~\cite{Topoto_weyl_1,Topoto_weyl_2}, implying the capability of protecting physical fields. Inspired by these, we may question whether we can extend the protection mechanism into quantum regime to directly protect quantum-correlated states. Together with the integrability and controllability of integrated photonic chip~\cite{g2_QW,g2_wg}, the on-chip topological boundary states may provide an alternative way of protecting quantumness effectively.

In this letter, we experimentally investigate the evolution dynamics and their preservation of two-photon quantum correlated states in the topological photonic lattice on a photonic chip. We observe no distinct drop of cross-correlation in boundary state. In contrast, the two photons in bulk states are found to dissipate into many sites and a few best measurable sites only give a cross-correlation of 6 with very large uncertainty, with which a quantum entanglement derived will not be able to violate Bell inequality any more~\cite{Bell_1,Bell_2}. We also verify the preservation of two-photon quantum correlation in the boundary state by measuring a strong violation of Cauchy-Schwarz inequality up to 30 standard deviations~\cite{CS}. 

\begin{figure*}
	\centering
	\includegraphics[width=1.8\columnwidth]{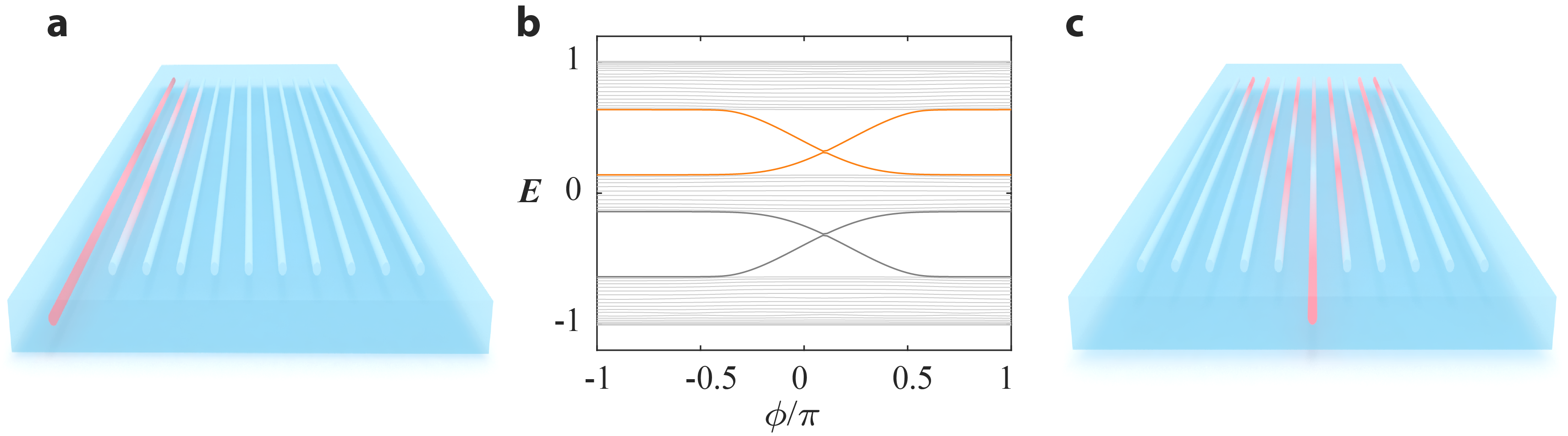}\\
	\caption{\textbf{Band structure and the schematic diagram of the topological system.} The photons will be confined in the boundary of the lattice when the photons are injected from the boundary with $\phi =0.2$ (a), as the result of that the gaps between bands are crossed by two topologically nontrivial boundary modes in band structure (b), and the photons injected from the middle site tend to evolve into different sites for the bulk state (c).}
	\label{f1}
\end{figure*}

\begin{figure*}
	\centering
	\includegraphics[width=1.8\columnwidth]{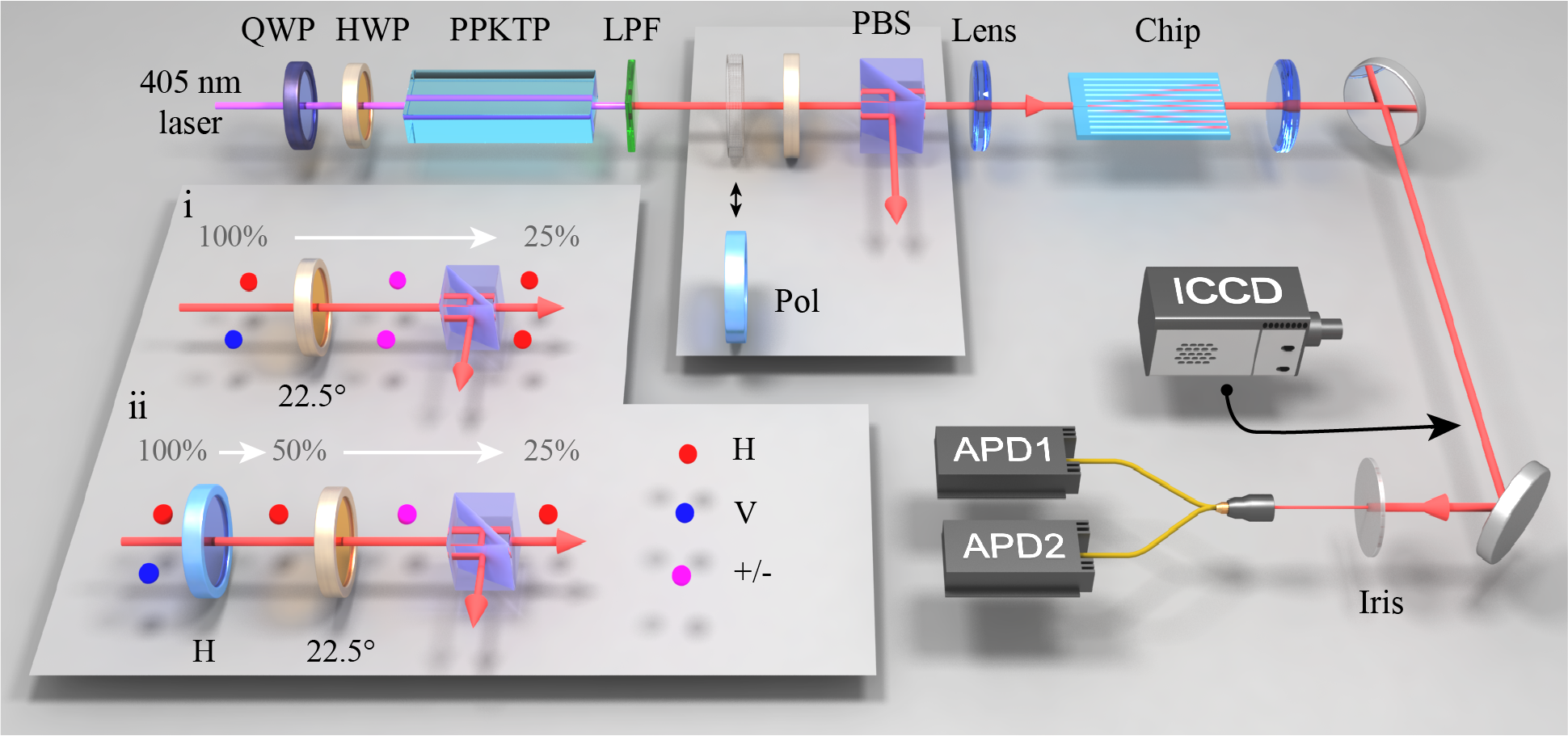}\\
	\caption{\textbf{Experimental setup.} Pair-like correlated photons with a wavelength of 810 nm are simultaneously generated from PPKTP crystal, both photons are prepared to horizontal polarization after a half wave plate (HWP) and a polarizing beam splitter (PBS) with 25\% probability, as is shown in the inset \textbf{i}, and then are injected into the lattice. Two photons out of certain site, spatially selected by an iris, are detected by APDs respectively after being split by a fiber beam splitter. A counter (not shown in setup) is used to record the coincidence of the photon pairs. The single counts and coincidence together can give the cross-correlation. By inserting a polarizer (Pol), also shown in the inset \textbf{ii}, the signal photon and the idler photon can be chosen individually with the same probability of 25\%, and then their auto-correlation can be measured separately. The detection can be switched into a single-photon sensitive ICCD for the measurement on photon outgoing distribution. QWP: quarter wave plate, LPF: long-pass filter.}
	\label{f2}
\end{figure*}

\begin{figure*}[!t]
	\centering
	\includegraphics[width=1.4\columnwidth]{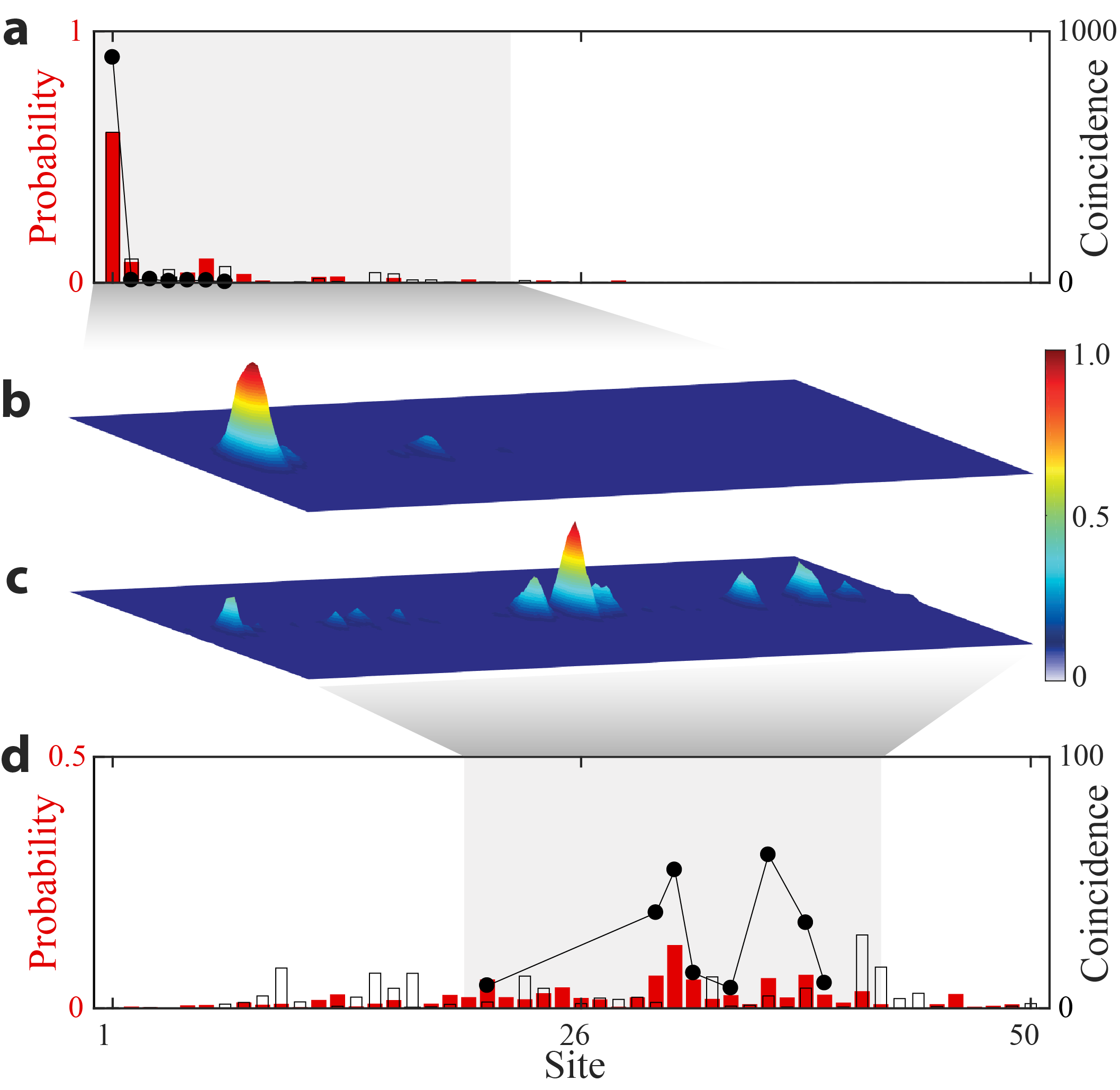}\\
	\caption{\textbf{Experimental results of the photon probability distribution and the coincidence counts (labeled in right) over 300 seconds.} \textbf{a,} The measured probability distribution of outgoing photons from the boundary state (represented with red histogram), and the measured coincidence counts over 300 seconds at the output of the lattice (represented with black dots). \textbf{b, c,} Accumulated part images of photon probability distribution over 1500 seconds of boundary state (b) and bulk state (d) obtained from ICCD. The colorbar normalizes the distribution with the maximal value of each sample. \textbf{d,} The measured results for the outgoing photons from the bulk state. The measured coincidence counts are much smaller than that in the boundary state.}
	\label{f3}
\end{figure*}

We construct a topological boundary state under an \textit{off-diagonal} Harper model (also known as Aubry-Andr{\'e} model)~\cite{AA1,AA2}. We can describe the system by the Hamiltonian
\begin{equation}
    \label{eq1}
     H=\sum_{n}t(1+\lambda cos(2\pi bn+\phi))a_na_{n+1}^++h.c.,
\end{equation}
where $a^+_n$ is the creation operator at site $n$, $t$ represents the average coupling strength between the adjacent sites, $\lambda$ is the modulation amplitude, $b$ is the periodic parameter and $\phi$ is the initial phase. Such a system inherits its robust boundary states from the two-dimension integer quantum Hall effect and possesses similar topological characteristics~\cite{QC1}.

We set the modulation amplitude $\lambda =0.5$ and the site number $N=50$. The calculated Floquet band structure as a function of phase $\phi$ is shown in Fig.\ref{f1}. We can see two topologically nontrivial boundary modes in the gap connecting different bands. Under the restraint of the boundary state, the photons are confined in the boundary of the lattice when the system is excited in the leftmost site  (n=1). As a comparison, for the bulk state, the photons are no longer confined in one or two sites and spread to other sites as a typical quantum walk in quasi-crystal.

We fabricate the lattice in a borosilicate glass using femtosecond laser direct writing technique~\cite{FH,fabri_1,fabri_2,fabri_3,fabri_4}(see Methods) with phase $\phi =0.2\pi$, where there exists two topologically nontrivial boundary modes as shown in Fig.\ref{f1}. The quantum-correlated photon pairs are injected into the leftmost $(n=1)$ and the middle $(n=26)$ sites respectively. The outgoing probability distribution and the cross-correlation quality are measured after the photons have evolved for 35 mm.

The experimental setup in Fig.\ref{f2} shows that a horizontally and a vertically polarized photons with a wavelength of 810 nm are simultaneously generated from PPKTP crystals via spontaneous parametric down conversion. Both of them are transformed to horizontal polarization with 25\% probability after a designed polarization rotation and projection. Then they are injected into the lattice simultaneously. We measure the photon outgoing distribution with a single-photon sensitive intensified CCD (ICCD) camera. We detect the cross-correlation of the photon pairs by avalanche photodiodes (APDs) after a fiber beam splitter. More details about quantum light source and experimental measurement can be found in Methods.

The outgoing photon probability distribution of the boundary state is shown in Fig.\ref{f3}(a-b), where the photon can almost only be found in the leftmost site. The coincidence of correlated photon pairs detected over 300 seconds at the output of the left seven sites is shown by the black dots. As we can see, the two photons simultaneously occupy the leftmost site with a high probability up to 94.6\%. Meanwhile, when we inject the quantum-correlated photon pair in the middle site, a trivial quantum walk type scenario for the bulk state is observed. As is shown in Fig.\ref{f3}(c-d), the two photons spread across the lattice, so that even the coincidences in a few best measurable sites are still much smaller than that in the topological boundary state.

\begin{figure*}[!t]
	\centering
	\includegraphics[width=1.3\columnwidth]{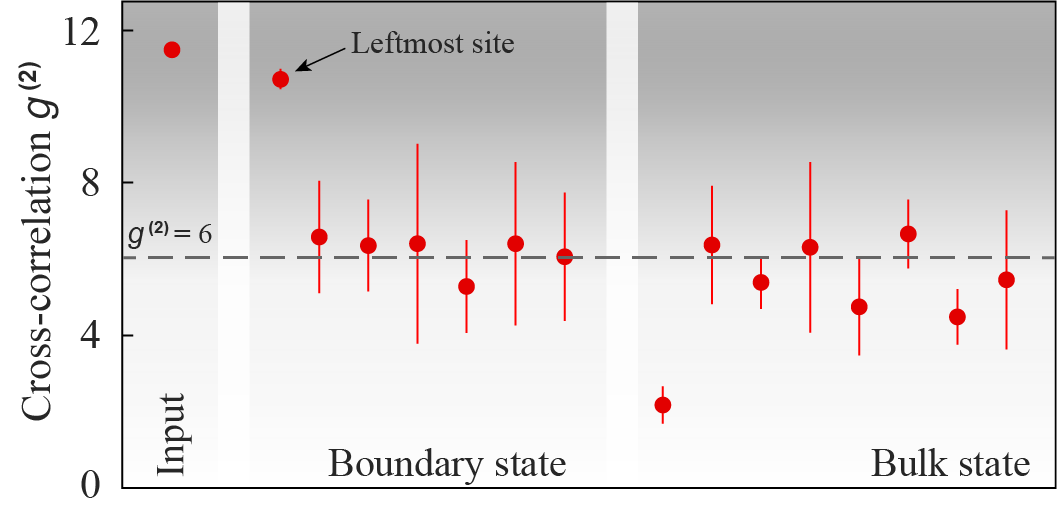}\\
	\caption{\textbf{Measured cross-correlation $g^{(2)}$.} The value of $g^{(2)}_{s\text{-}i}$ for the leftmost site (n=1) up to $10.70\pm 0.25$ approaches to that of the input state. The sites near by the leftmost site for the topologically nontrivial boundary state and the sites in the bulk sate all have comparably much smaller $g^{(2)}_{s\text{-}i}$ as well as very large variances.}
	\label{f4}
\end{figure*}

\begin{table*}
    \centering
	\par    
	\caption{\textbf{The measured cross-correlation, auto-correlation and the violation of Cauchy-Schwarz inequality.} The $Input$ means the source before chip, $Boundary$ represents the leftmost site in boundary state, and $Bulk_i$ is the site $i$ in bulk state. We denote the standard deviations of violation of Cauchy-Schwarz with $SDs$ }
	
	\begin{tabular}{p{1.8cm}<{\centering} p{1.8cm}<{\centering} p{1.8cm}<{\centering} p{1.8cm}<{\centering} p{1.8cm}<{\centering} p{1.8cm}<{\centering} p{1.8cm}<{\centering}}
		\hline\noalign{\smallskip}
		\hline\noalign{\smallskip}
		
		$Quantity$          & $Input$    &  $Boundary$   & $Bulk_{30}$  & $Bulk_{31}$ & $Bulk_{36}$   & $Bulk_{38}$   \\[0.4cm]
		
		$g^{(2)}_{s\text{-}i}$ & 11.47(2)     & 10.70(25)     & 6.37(154)    &  5.39(69)    &  6.66(90)     &  4.49(73)     \\[0.25cm]
		
		$g^{(2)}_{s\text{-}s}$ & 2.00(2)       & 1.79(17)       & 3.90(87)   &  1.72(86)    &  1.57(47)     &  1.71(47)      \\[0.25cm]
		
		$g^{(2)}_{i\text{-}i}$  & 1.82(1)       & 1.67(50)       & 2.36(63)   &  3.01(174)    &  1.79(60)     &  2.97(121)     \\[0.25cm]
		
		$SDs$                         & 410.42         & 30.48          & -1.84        &  0.31            &  4.29            &  -0.79     \\[0.1cm]
		
		\hline\noalign{\smallskip}
		\hline\noalign{\smallskip}	
	\end{tabular}   
\label{tab1}
\end{table*}

To quantify the performance on preserving the two-photon quantum correlation, we measure the cross-correlation function, $g^{(2)}_{s\text{-}i} = p_{s\text{-}i}/p_sp_i$~\cite{Q_optics}, before and after our photonic chip. The value of $g^{(2)}_{s\text{-}i}$ for the leftmost site  in the boundary state is up to $34.12\pm 1.44$, which is still comparable to $44.31\pm 0.16$ obtained before the photonic chip. For the bulk state, the highest value of $g^{(2)}_{s\text{-}i}$ we can get is $15.12\pm 2.14$, and the values for the other sites are very difficult to obtain due to the low coincidence as aforementioned. We increase the pump power to generate the photon pairs with a higher flux, which makes it possible to measure the cross-correlation (and also the auto-correlation, which will present later) in the bulk state in reasonable time.

As is shown in Fig.\ref{f4}, we measure more details under different pump condition. The cross-correlation of the protected state is $10.70\pm 0.25$, apparently very close to the value of $11.47\pm 0.02$ before the photonic chip. That means the boundary state preserves not only the photon probability but also their quantum correlation. As a comparison, we measure the cross-correlation for several sites near the leftmost site in the boundary state and for eight sites with the highest photon probability in the bulk state. All the measured values are found to be as low as 6 with very large uncertainty, with which a quantum entanglement derived will not be able to violate Bell inequality any more~\cite{Bell_1,Bell_2}. It implies that the quantum features of the two-photon states tend to degenerate without the protection of topological boundary state.

As a way of distinguishing from classical behaviors, the non-classicallity can be further revealed clearly by a violation of Cauchy-Schwarz inequality \cite{CS}
\begin{equation}\label{eq2}
{\left({g_{s\text{-}i}^{\left( 2 \right)}} \right)^2} \le g_{s\text{-}s}^{\left( 2 \right)} \cdot g_{i\text{-}i}^{\left( 2 \right)}
\end{equation}
where $g^{(2)}_{s\text{-}s}$ ($g^{(2)}_{i\text{-}i}$) is the auto-correlation of the signal (idler) photon. As is shown in Table.\ref{tab1}, we measure the inequality for the input state, the protected state, and the states outgoing from the four best measurable sites in the bulk state. Again, we can observe a clear violation of Cauchy-Schwarz inequality for the protected state but not for the unprotected state.

In summary, we experimentally demonstrate topological protection of two-photon quantum correlation on a photonic chip. The measurement on the cross-correlation and the violation of Cauchy-Schwarz inequality indicate that the two-photon quantum-correlated states can be well preserved in the topologically nontrivial boundary state, and substantially thermalize in the bulk state. Our results extend the protection mechanism of topological phases into quantum regime to directly protect quantum-correlated states, representing an emerging and alternative way of protecting quantumness. Further works include the protection of quantum entanglement or the implementation in higher dimensional structures.

\subsection*{Acknowledgments}
The authors thank Jian-Wei Pan for helpful discussions. This work was supported by National Key R\&D Program of China (2017YFA0303700); National Natural Science Foundation of China (NSFC) (11374211, 61734005, 11690033); Shanghai Municipal Education Commission (SMEC)(16SG09, 2017-01-07-00-02-E00049); Science and Technology Commission of Shanghai Municipality (STCSM) (15QA1402200, 16JC1400405). X.-M.J. acknowledges support from the National Young 1000 Talents Plan.

\subsection*{Methods}
{\bf Fabrication of the photonic lattice:} We design the lattice according to the characterized coupling coefficient, which is modulated by the separation between two adjacent waveguides. The waveguides and the whole lattice are written in borosilicate glass (refractive index $n_0=1.514$) by using a femtosecond laser with the working wavelength 513nm, repetition rate 1MHz and pulse duration 290fs. We reshape the laser writing beam with a cylindrical lens, and then focus the beam inside the borosilicate substrate with a 50X objective lens (NA 0.55),  We move the substrate during fabrication with a constant velocity of 10mm/s using a high-precision three-axis motion stage.\\

{\bf Generation and measurement of the correlated photons:} The pair-like photons are generated by pumping a periodically-poled KTP (PPKTP) crystal via spontaneous parametric down conversion process. With designed polarization rotation and projection, as shown in the inset of Fig.\ref{f2}, we can prepare the correlated photons in identical polarization with a probability of 25\%. To measure the cross-correlation, the photons out of the lattice are detected with APDs and a FPGA counter after being split by a fiber beam splitter. It should be noticed that the photon flux injected into the lattice is not fair with the case of two photons if we just move the half wave plate to measure the auto-correlation $g_{s\text{-}s}$($g_{i\text{-}i}$) of signal (idler) photon. We add one polarizer before the half wave plate to make sure the photon flux the same for all measurement scenarios. To measure the outgoing photons, we inject them into the lattice using a 20X objective lens, and observe the evolution results from the lattice using a 10X microscope objective lens.

\end{document}